\documentclass[twocolumn,twoside,slac_two]{revtex4}
\usepackage{graphicx}
\usepackage{fancyhdr}
\begin{document}
\title{Observation of the Perseus cluster of galaxies with the MAGIC telescopes}
\author{Saverio Lombardi}
\affiliation{Universit\`{a} di Padova and INFN, I-35131 Padova, Italy}
%
%
\author{Pierre Colin}
\affiliation{Max-Planck-Institut f\"{u}r Physik, D-80805 M\"{u}nchen, Germany}
%
%
%
\author{Dorothee Hildebrand}
\affiliation{ETH Zurich, 8093 Zurich, Switzerland}
%
%
%
%
%
\author{Fabio Zandanel, Francisco Prada}
\affiliation{Instituto de Astrof\'{\i}sica de Andaluc\'{\i}a (CSIC), E-18080 Granada, Spain}
\author{\emph{on behalf of the MAGIC Collaboration}}
\affiliation{http://magic.mppmu.mpg.de/}
%
%
\author{Christoph Pfrommer}
\affiliation{Heidelberg Institute for Theoretical Studies (HITS), D-69118 Heidelberg, Germany}
\author{Anders Pinzke}
\affiliation{UCSB, California, USA}
%
%
%
\begin{abstract}
The MAGIC telescopes performed a deep observation of the central 
region of the Perseus galaxy cluster in stereoscopic mode between 
October 2009 and February 2011. The nearly 85~hr of collected data 
(after quality selection) represent the deepest observation of a cluster 
of galaxies at very high energies (VHE, $E>100$~GeV) ever. The 
survey resulted in the detection of VHE $\gamma$-ray emissions from 
its central galaxy NGC~1275 and from the radio galaxy IC~310. 
In addition, the deep survey also permits for the first time to 
constrain emission models predicting VHE $\gamma$-rays from 
cosmic-ray acceleration in the cluster. In this contribution we 
report the latest MAGIC results concerning these topics.
\end{abstract}
\maketitle
\thispagestyle{fancy}
\section{Introduction}
Clusters of galaxies are the latest objects to form in the Universe 
and represent the largest and most massive gravitationally bound systems, 
with radii of few Mpc and total masses $M \sim (10^{14}-10^{15})M_{\odot}$, 
of which galaxies, gas, and dark matter (DM) contribute roughly for 
5$\%$, 15$\%$, and 80$\%$ respectively (see e.g.~\cite{voit2005} 
for a general overview).\\
Clusters of galaxies are very complex astrophysical environments, where 
a significant amount of very high energy (VHE, $E>100$~GeV) $\gamma$-ray 
emission is expected on the following general grounds.
(i) Clusters are actively evolving objects and they should dissipate 
energies of the order of the final gas binding energy through merger 
and accretion shocks as well as turbulences, which are also likely to 
accelerate non-thermal electrons and protons to high energies~\cite{pfrommer2008}. 
(ii) Clusters are home of different types of energetic outflows of 
powerful sources such as radio galaxies~\cite{forman2003} and  
supernova-driven galactic winds~\cite{volk2003}.
(iii) Clusters contain large amounts of gas with embedded magnetic 
fields often showing direct evidence for shocks and turbulence as 
well as relativistic particles~\cite{feretti2003}.\\
Furthermore, galaxy clusters are characterized by very large 
mass--to--light ratios and considerable DM overdensities, and hence they 
could be considered as interesting targets for the search of emissions 
in the $\gamma$-ray regime from DM annihilation~\cite{masc2011,pinzke2011} 
or decay~\cite{cuesta2011}.
However, the recently underlined very extended nature of the DM signal in 
clusters~\cite{masc2011,pinzke2011} represents a major issue for the current 
generation of Cherenkov telescopes.\\
The Perseus cluster, at a distance of 77.7~Mpc (z~=~0.018), is the 
brightest X-ray cluster~\cite{rei02}, hosting a massive cooling flow 
and a luminous radio mini-halo that fills a large fraction of the cluster 
core region~\cite{pedlar1990}. The radio mini-halo is well explained by 
the hadronic scenario where the radio emitting electrons are produced in 
hadronic CR proton-proton interactions with intra-cluster medium (ICM) protons~\cite{pfrommer2004}.
Additionally, the Perseus cluster hosts two very interesting objects, 
that have been recently detected at high energy (HE, $100$~MeV~$<E<$~$100$~GeV) 
by \emph{Fermi}--LAT, and at VHE by MAGIC, namely the two radio galaxies
NGC~1275 and IC~310.\\
NGC~1275 is the central galaxy of the Perseus cluster and is of great 
interest due to its possible ``feed-back'' role in the cluster environment 
(e.g.~\cite{gallagher2009}), and for physics studies of relativistic outflows.
Its classification varies between different papers and catalogues, and 
the complex structure of NGC 1275 including surrounding filaments 
leads to a peculiar morphology classification~\cite{dev1991}. \\
IC~310 is classified as head-tail radio galaxy, a type of active galactic 
nuclei only occuring in dense galaxy clusters~\cite{begelman1979}, like the Perseus cluster.
In 1999, it was suggested that IC~310 could be a dim blazar because of the absence 
of strong emission lines and the spectral indices on radio and X-ray 
measurements~\cite{rector1999}. Later on it was also shown that the X-ray 
emission may originate from the central active galactic nucleus
of a BL Lac-type object~\cite{sato2005}.
\section{The MAGIC Telescopes and the observations of the Perseus Cluster}
\begin{figure}
\centering
\includegraphics[width=80mm]{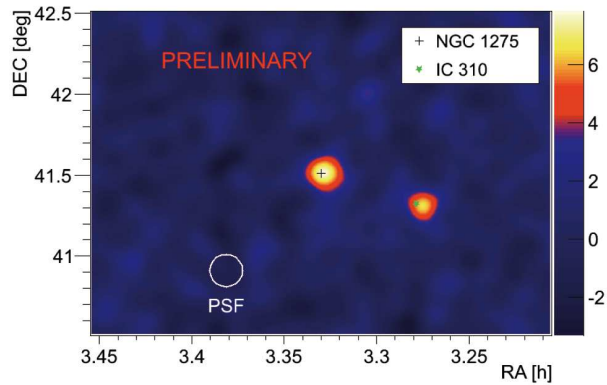}
\caption{
Significance skymap above $150$~GeV of the Perseus cluster region.
For this map the overall MAGIC stereo data (nearly $85$~hr, after selection)
between October 2009 and February 2011 have been used. NGC~1275 is clearly detected 
at the center of the cluster. The second significant excess corresponds 
to IC~310, $0.6^{\circ}$ away from the center of the cluster.
}
\label{skymap}
\end{figure}

The MAGIC experiment consists of two 17~m dish Imaging Air Cherenkov Telescopes 
(IACTs) located on the Canary Island of La Palma (2200~m a.s.l.). The MAGIC 
telescopes are currently the largest existing IACTs. Since the end of 2009 
the telescopes are working together in stereoscopic mode which ensures an 
excellent sensitivity of $0.8\%$ of Crab Nebula flux above $\sim 300$~GeV in 50~hr 
of observations, and a trigger energy threshold of $50$~GeV~\cite{aleksic2011a}. 
The angular and energy resolution at $100$~GeV are $0.1^{\circ}$ and $20\%$ 
respectively. The stereoscopic observations improved the sensitivity
achieved with the single telescope observations by a factor of $\sim$2 for 
energies above few hundreds of GeV, and a factor $\sim$3 for lower energies down 
to the threshold, which allows us to extend the observations carried out by 
the \emph{Fermi}--LAT detector up to the TeV scale and without energy gaps.\\
The Perseus galaxy cluster was carefully chosen over other nearby clusters as 
it is the most promising target for the detection of $\gamma$-rays originating from 
the neutral pion decays result of the hadronic cosmic-rays (CR) interactions 
with the ICM~\cite{pinzke2010}. 
Additionally, the central radio galaxy NGC~1275 is a very promising GeV-TeV target~\cite{Fermi_1,Fermi_2,Fermi_3}. 
Hence it represents \emph{per se} a good reason for the observation of this cluster at VHE.\\
The MAGIC experiment conducted the deepest survey ever made at VHE of the Perseus 
cluster, collecting data in both single telescope mode ($\sim$25~hr of MAGIC-I 
observations between November and December 2008)~\cite{aleksic2010a} and stereoscopic mode 
($\sim$85~hr of observations between October 2009 and February 2011)~\cite{aleksic2011b,aleksic2011c}. 
The source was observed in the false source tracking (wobble) mode~\cite{fomin1994}, 
with data equally split in different pointing positions located symmetrically 
at $0.4^\circ$ from NGC~1275, in order to ensure optimum sky coverage and 
background estimate. The survey was carried out during dark time at low zenith 
angles (from $12^\circ$ to $36^\circ$), which guaranteed the lowest energy threshold. 
The analysis of the data was performed using the standard MAGIC software~\cite{aliu2009}, 
taking advantage (for the stereoscopic data) of newly developed analysis routines~\cite{aleksic2011a,lombardi2011}.
These observations resulted in the discovery of IC~310~\cite{aleksic2010b} and NGC~1275 
(ATel$\#$2916)~\cite{aleksic2011c} as VHE $\gamma$-ray emitters, as shown in figure~\ref{skymap}. 
%
\section{MAGIC-I observation of the Perseus cluster}
\begin{figure}
\centering
\includegraphics[width=75mm]{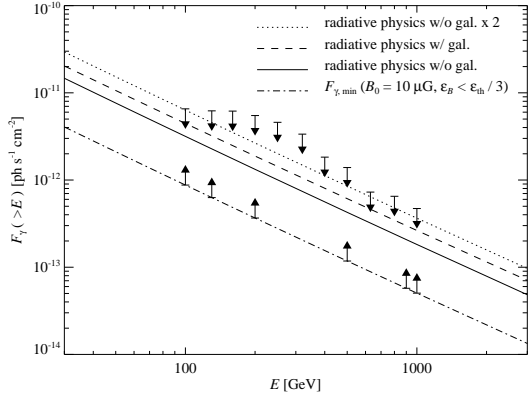}
\caption{
Comparison between the MAGIC-I (mono) observation integral flux upper limits (upper arrows) and the simulated 
integrated spectra of the CR induced $\gamma$-ray emission of the Perseus cluster~\cite{pinzke2010}. 
The conservative model without galaxies (solid) is contrasted to the model with galaxies (dashed) and it is 
scaled by a factor of 2 so that it is just consistent with the upper limits obtained in~\cite{aleksic2010a} (dotted). 
Additionally shown is the minimum $\gamma$-ray flux estimated for the hadronic model of the radio mini-halo of 
the Perseus cluster (dash-dotted with arrows). 
Image taken from~\cite{aleksic2010a}.
}
\label{perseusULs}
\end{figure}

The Perseus galaxy cluster was observed by the MAGIC-I telescope for a total
observation time (after data selection) of 24.4~hr during November~--~December 2008. 
No significant excess was found in the data~\cite{aleksic2010a}. The integral flux 
upper limits (at $95\%$ confidence level)
were compared to the simulated flux of the $\gamma$-ray emission from decaying neutral pions that 
result from hadronic CR interactions with the ambient gas in the Perseus cluster~\cite{pfrommer2004}, 
allowing to constrain the average CR-to-thermal pressure to $<4\%$ for the cluster 
core region and to $<8\%$ for the entire cluster (see figure~\ref{perseusULs}).
\section{Stereoscopic observations of the Perseus cluster}
The Perseus galaxy cluster region was observed by the MAGIC telescopes
(in stereoscopic mode) during two distinct campaigns. The first one was
carried out between October 2009 and February 2010, for a total observation
time of $45.3$~hr. This survey resulted in the discovery of the radio galaxy
IC~310 as VHE emitter~\cite{aleksic2010b}. The latest campaign (total observation time
of $53.6$~hr), which resulted in the discovery of NGC~1275 at VHE (ATel$\#$2916)~\cite{aleksic2011c},
was performed between August 2010 and February 2011.\\ 
The whole stereoscopic data sample ($\sim$85~hr after data selection) was
used to investigate a possible signal from CR hadronic interactions. 
For this purpose, the analysis was restricted to energies where the central 
radio galaxy NGC~1275 is not emitting~\cite{aleksic2011b}, i.e. approximately 
above $600$~GeV.
No CR induced emission is detected above those energies, and the preliminary 
integral flux upper limit above $1$~TeV (at 95$\%$ confidence level)
are 
about a factor $\sim$3 more constraining that those achieved with the
MAGIC-I mono observation. This permits to significantly tighten the previous 
constraints, and to start to probe the acceleration physics of CR at structure 
formation shocks. The estimation and interpretation of the flux upper limits are 
ongoing and the corresponding paper is under preparation~\cite{aleksic2011c}.\\
\subsection{Discovery of VHE emission from IC~310}
\begin{figure}
\centering
\includegraphics[width=70mm]{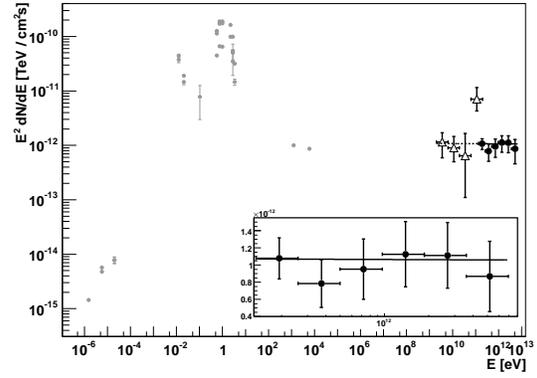}
\caption{
Spectral energy distribution of IC~310 obtained with $20.6$~hr of MAGIC stereo data 
(full circles). The open triangles show the flux measurements from the first two years 
of operation of \emph{Fermi}--LAT. Archival X-ray, optical, IR, and radio data obtained 
from the NED database are shown with grey dots. The solid line shows a power-law fit to 
the MAGIC data, and the dotted line is its extrapolation to the GeV energies. 
A zoom-in of the MAGIC points is also shown. Image taken from~\cite{aleksic2010b}.
}
\label{IC310_SED}
\end{figure}

\begin{figure}
\centering
\includegraphics[width=70mm]{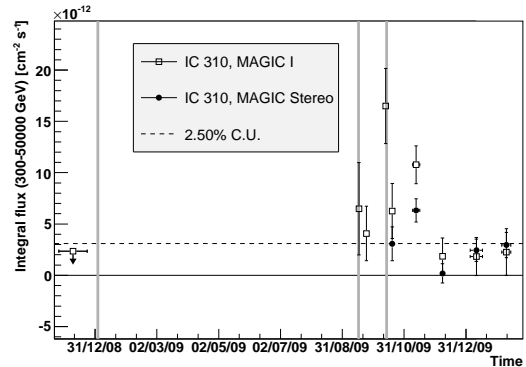}
\caption{
Light curve (in 10-day bins) of the $\gamma$-ray emission above $300$~GeV obtained with
the mono (black) and the stereo (red) MAGIC data. The black square with an arrow is 
the upper limit on the emission in November~--~December 2008. Vertical grey lines show 
the arrival times of photons above $100$~GeV from the \emph{Fermi}--LAT instrument. 
The horizontal dashed line represents a flux level of 2.5$\%$ Crab Nebula flux.
Image taken from~\cite{aleksic2010b}.
}
\label{IC310_LC}
\end{figure}

The radio galaxy IC~310 (redshift z=0.019) is located at a distance of $0.6^{\circ}$ 
from the cluster's central galaxy, NGC~1275. The source has been discovered in 2010 
by the \emph{Fermi}--LAT detector at HE~\cite{neronov2010} and by MAGIC at VHE~\cite{aleksic2010b}.
The combined MAGIC and \emph{Fermi}--LAT spectrum is consistent with a flat 
spectral energy distribution stretching without a break over more than 3 orders 
of magnitude in energy ($2$~GeV~–-~$7$~TeV), as shown in figure~\ref{IC310_SED}. 
The spectrum at VHE measured by MAGIC has a spectral index of $\Gamma$~=~$-2.00\pm0.14$,
and the mean flux above $300$~GeV, between October 2009 and February 2010, is
$F_{\gamma}$~=~$(3.1\pm0.5)\times10^{-12}~\mathrm{cm^{-2}~s^{-1}}$.
Hints of week to year time-scale variability were seen in the MAGIC data 
(see figure~\ref{IC310_LC}).\\
\subsection{Discovery of VHE emission from NGC~1275}
\label{NGC1275discovery}
\begin{figure}
\centering
\includegraphics[width=75mm]{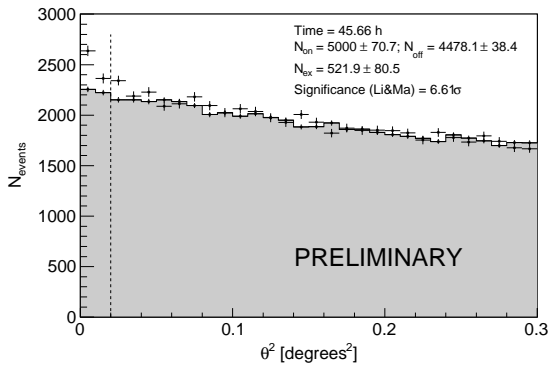}
\caption{$\theta^2$ distribution of the signal and the background
estimate from $45.7$~hr of MAGIC stereo observations (after data selection)
taken between August 2010 and February 2011, above $100$~GeV.
The region between zero and the vertical dashed line represents the
so-called signal region, within which the excess event number is estimated.}
\label{theta2}
\end{figure}

The central cluster radio galaxy NGC~1275 was first detected in the HE $\gamma$-ray regime 
by the \emph{Fermi}--LAT detector~\cite{Fermi_1}, during the first 
four months of \emph{all-sky-survey} observations, with a differential energy spectrum 
between $100$~MeV and $25$~GeV well described by a power-law with spectral index 
of $\Gamma$~=~$-2.17\pm0.05$, and an average flux above $100$~MeV of 
$F_{\gamma}$~=~$(2.10\pm0.23)\times10^{-7}~\mathrm{cm^{-2}~s^{-1}}$.
Subsequent \emph{Fermi}--LAT observations~\cite{Fermi_2,Fermi_3} have revealed that the 
average $\gamma$-ray spectrum of NGC~1275 shows a significant deviation from a simple power-law. 
The observed \emph{Fermi}--LAT spectrum is fitted best by a power-law function ($\Gamma \simeq -2.1$) 
with an exponential cut-off at the break photon energy of tens of GeV. This spectral behavior 
is compatible with the upper limits on the flux at VHE provided by MAGIC-I~\cite{aleksic2010a} 
and VERITAS~\cite{acciari2009}.\\
The source has been detected for the first time above $100$~GeV by the MAGIC telescopes
in the data taken between August 2010 and February 2011. Figure~\ref{theta2} shows 
the $\theta^2$ distribution (the $\theta^2$ being the squared angular distance between the arrival 
direction of the events and the nominal source position~\cite{daum1997}) of the signal coming 
from NGC~1275 and of the background (estimated from 3 distinct regions), for energies above $100$~GeV.
An excess of $521.9\pm80.5$ events, corresponding to a $6.6$$\sigma$ significance 
(calculated according to the equation~17 in~\cite{lm1983}) was found. The observed flux 
is estimated to be $\sim$2.5$\%$ of the Crab Nebula flux above $100$~GeV, and it decreases 
rapidly with energy. No signal is detected approximately above $600$~GeV.
A dedicated paper concerning this discovery is in preparation~\cite{aleksic2011b}. 
%
%
\section{Conclusions}
We presented the results achieved so far from the deep survey of the Perseus cluster of galaxies at VHE
carried out by the MAGIC experiment, both in mono ($\sim$25~hr) and stereo ($\sim$85~hr) data taking mode.
Mono observations permitted to constrain the average CR-to-thermal pressure to $<4\%$ for the cluster
core region and to $<8\%$ for the entire cluster~\cite{aleksic2010a}. 
The implications of the significantly tighter upper limits from the stereo observations are currently under 
investigation and will be published in a forthcoming paper~\cite{aleksic2011c}.\\
The stereoscopic observation of the Perseus cluster resulted on the discovery of two VHE $\gamma$-ray sources:
the radio galaxy IC~310~\cite{aleksic2010b}, which is characterized by a flat spectrum in the GeV to TeV range 
with hints of year to week time-scale variability, as well as the central cluster radio galaxy NGC~1275 (ATel$\#$2916)~\cite{aleksic2011b}
which has been detected below $600$~GeV, thanks to the excellent sensitivity of the MAGIC stereoscopic 
system in that energy range.
%
%
\bigskip 
%
%

%
%
\end{document}